\documentclass[preprint,onecolumn,10pt]{elsarticle}
\usepackage{amssymb}
\usepackage{amsfonts}
\usepackage{times}
\usepackage{graphicx}
\usepackage{float}
\usepackage{indentfirst}
\usepackage{amsmath}
\usepackage{epstopdf}
\usepackage{color}
\usepackage{textcomp}
\usepackage{subfigure}
\usepackage[all]{xy}
\usepackage{longtable}
\usepackage{amssymb}
\usepackage{lineno,hyperref}
\modulolinenumbers[5]

\journal{Journal of \LaTeX\ Templates}

\begin{document}

\begin{frontmatter}
\title{Epidemic spreading dynamics with drug-resistant and heterogeneous contacts\tnoteref{label1}}
%\tnotetext[label1]{}
\author[label1,label2]{Peng-Bi Cui\corref{cor2}}
\author[label1,label2]{Lei Gao\corref{cor3}}
\author[label3]{Wei Wang\corref{cor1}}
\ead{wwzqbx@hotmail.com}
\cortext[cor1]{Corresponding author at: College of Computer Science and Technology, Chongqing University of Posts and Telecommunications, Chongqing 400065, China}
\address[label1]{Web Sciences Center, University of Electronic Science and Technology of China, Chengdu 611731, China}
\address[label2]{Big Data Research Center, Univesrsity of Electronic Science and Technology of China, Chengdu 611731, China}
\address[label3]{College of Computer Science and Technology, Chongqing University of Posts and Telecommunications, Chongqing 400065, China}

%\fntext[label3]{}

\title{Epidemic spreading dynamics with drug-resistant and heterogeneous contacts}

\begin{abstract}
Drug resistance and strong contacts actually play crucial
roles in epidemic spread in complex systems. Nevertheless,
neither theoretical model or methodology is proposed to address
this. We thus consider an edge-based epidemic spread model
considering the two key ingredients, in which the contacts
are grouped into two classes: strong contacts and normal ones.
Next, we present a unified edge-based compartmental approach to
the spread dynamics on Erd\"{o}s-R\'{e}nyi (ER) networks and
validate its results by extensive numerical simulations.
In case that epidemic is totally drug-resistant, we both numerically
and theoretically show a slow outbreak (continuous transition) of
epidemics when number of strong contacts is not enough for the
emergence of null threshold. If the epidemic owns partial
resistance, we would observe evident faster-growing outbreaks
(discontinuous transitions) and larger final epidemic sizes for
few strong contacts, instead of emergence of null threshold with
increase of strong contacts. Inhibiting effect of infection
threshold, positive roles of strong contacts and strength of
strong contacts in promoting outbreaks are also approved.
Throughout this paper, we could drive exact predictions
through the analytical approach, showing good agreements
with numerical simulations.

\end{abstract}


\end{frontmatter}

%%\linenumbers

\section{Introduction}
\label{sec:introduction}
In field of epidemic research, an actual non-negligible case that is recently gaining attention is drug resistance arisen from abuse of substance especially antibiotics, which contributes to the problem of drug resistance. Drug resistance refers to that individually applied therapies targeted at single pathogens in individual bodies would actually become environmental events to drive the evolution of pathogens and commensal bacteria alike far beyond bodies. Growing problems with multi-resistant pathogens such as methicillin-resistant staphylococcus aureus (MRSA) found in most European countries~\cite{mrsa}, more particularly infectious agents of tuberculosis (like multidrug-resistant tuberculosis (TB) or even extensively drug-resistant TB~\cite{totally}) and gonorrhoea begin to capture much more attention than before, because these pathogens remaining high infection rates would make infected people are difficult or impossible to cure. For some extreme cases of multi-drug resistant epidemics like superbugs, even last-resort antibiotics fail~\cite{chen2017,totallyindia}. Therefore large amount of empirical studies have been conducted by epidemiologists, microbiologists, health economists and physicians to the drug resistance especially antibiotic resistance, aiming to develop new medicines, therapies even coping strategies in the face of this crisis~\cite{antibiotic1}.

In turning to studies related to epidemic spread in complex systems, lots of researchers have developed different models, estimation methods or efficient algorithms to explore the influence of spatial structure topology especially the edges on spread of epidemics, so as to design network-based prevention strategies~\cite{ccrmp}. However, designing network-based prevention strategies requires knowledge on disease transmission through network edges and statistical methods for analyzing network data. Fortunately, data-driven method such as contact tracing is very effective, because this method could identify which contacts are key to the transmission and can contribute to the so-called real time tracking of an epidemic~\cite{tracing}. Further, one significant research progress of contact tracing is that there exist considerable differences among contact frequencies in real complex networks~\cite{tracing,contactfrequency}, while the frequency of contact can be used as a proxy measure for the tendency of at-risk events for infection. This means that various contact frequencies lead to multiple transmission rates (positively related to contact frequency) along different contacts. Hence, contact-based network models accessing nature of contact patterns could yielded more deep insights into infectious disease transmission, and to control their epidemiological significance.

Another important fact we could not ignore is that spread of epidemics with drug resistance is often accompanied by increasing number of some contacts with enough high strength (i.e. high contact frequency) which could be called strong contacts. And these strong contacts have been found to be widespread in reality, in forms of concurrent partnerships (spouse partnerships, cohabitation relationships and homosexual relations)~\cite{sexcontact1}, compact communities~\cite{tribes1,tribes2}, doctor-patient relationship~\cite{doctorpatient1,doctorpatient2} and so on. For example, gay relationship is always considered as a high risk contact for being infected with HIV compared with other forms of sexual contacts, through largely increasing infection rate~\cite{hiv1}. Another well-known case is that of tuberculosis transmissions, which could occur more easily among family members; despite the possible infection through inhalation of aerosols or small droplets for example during a conversation between strangers~\cite{tbfamily}. Besides, doctor-patient relationship has been confirmed as a nearly 'straightway unblocked channel' to transfer streptococci and staphylococci (even MRSA), resulting in a serious cross-infection~\cite{stap}. These studies indicate that the strength of contacts and drug resistances especially antimicrobial resistances can profoundly influence the spread of disease and should thus be incorporated into applicable epidemiological models. However, it still not clear how drug resistance regulates the epidemic diffusion in complex systems in presence of strong contacts. To our knowledge, no theoretical models have been proposed to center on this problem so far, to say nothing of a systemic methodology.

Based on the above arguments, we herein put forward an edge-based
epidemic spread model to capture the two realistic mechanisms:
drug resistance and strengths of contacts. In the present model,
contacts are directly classified into two types: normal contacts and strong ones, which allows us to develop a unified edge-based compartmental theory so as to deeply understand the effects of drug resistance and strong contacts on the spread dynamics. We elucidate the issue in two different situations at length: (1) Epidemic is totally drug-resistant; (2) Epidemic own partial resistance. In the first case, we both numerically and theoretically show a slow outbreak (continuous transition) of epidemics when number of strong edges (i.~e., strong contacts) is not sufficiently enough for the emergence of null threshold i.~e., the system exhibits the lack of an epidemic threshold and always show a finite fraction of infected population~\cite{nullthreshold}. We then extend the analytical approach to the latter case to derive a number of predictions, including the final infection size, the position of the threshold and the nature of the transition. Both the theoretical approach and agent-based simulations indicate the occurrences of faster-growing outbreaks (i.~e., discontinuous transitions) and larger final epidemic sizes with few strong contacts in networks, instead of emergence of null threshold with increasing fraction of strong contacts. In both cases, numerical simulations perfectly fit the analytical predictions.

The paper is organized as follows. In Sec.~\ref{sec:model}, we present a detailed description of our model. Then we give a detailed theoretical analysis of the spread dynamics by means of the unified edge-based compartmental theory in Sec~\ref{sec:theory}. Sec.~\ref{sec:results} is devoted to show a deep numerical investigation of the epidemic spread dynamics on different networks. In addition, theoretical predictions are given to give a comparison with numerical simulations. Finally, we list conclusions in Sec.~\ref{sec:conclusion}.
\section{Model}
\label{sec:model}

One propose a model to give a framework to curve spread of highly mutated epidemics which could avoid inhibitory effect of antibiotics, such as influenza virus, TB bacteria, staphylococci and so on. As we know, in the classical SIR model in discrete time, each individual can be in one of three different states: susceptible (S, individuals that are healthy, and could be infected), infected (I, individuals who get infected and can transmit the pathogen), or removed (R, dead or recovered and immunized individuals). At each time step, each infected individual shifts into removed state with probability $\gamma$, while she transmits the pathogen to each of her susceptible neighbors with infection probability $p$.

In the present model, the epidemic spread is formulated in the terms of
a modification of the susceptible-infected-recovered (SIR) model,
where one class the edges of networks into two groups: strong edges
and normal ones, corresponding to the strong contacts and normal contacts
which have been observed in reality. The reason that only two classes of
edges are assumed in the model is to enable us to develop a unified edge-based
compartmental approach to analysis the spread dynamics, without loss of
one important feature of human contacts -- heterogeneous contributions
to epidemic diffusion caused by different contact frequencies.
At the same time, we introduce infection threshold $T$ to depict
drug resistance of epidemics. Like what happens in the SAR model for social
contagions~\cite{weier1}, one susceptible individual successfully
becomes infected only if it has received the pathogen from its
connected infected neighbors for at least $T$ times (i. e.,
infection threshold). This infection rule with infection threshold
is actually a nice mirror to reflect how epidemics with drug
resistance diffuse in real complex systems~\cite{antibiotic1}.
The case $T=1$ represents one situation that the epidemic
is multi-resistant or even totally-resistant, so that it
could attacked people most severely, which means that individuals would
easily get infected if they receive pathogen of the epidemic from the
infected guys for just one time. For the case $T>1$, we know that
people have more choices in the antibiotic lists because the epidemic
owns partial resistance or even no. In this case, the epidemic could be
timely suppressed or repelled by acting antibiotics so that the hosts
could not infect others, until the pathogen begins to evolve certain
drug-resistance under the stress of antibiotics (receive times is beyond
infection threshold $T$).

Specifically, for two individuals $i$
(susceptible) and $j$ (recovered) located on two ends of one edge
$e_{ij}$ in the complex network, $i$ will receive the epidemic
pathogen passed from $j$ at a probability $p$ if the edge $e_{ij}$
is a normal edge. Instead, if $e_{ij}$ is a strong edge, $i$ will be
more likely to get infected by the epidemic transmitted from $j$ at
a raised  infection probability $q$. That is why our model is
edge-based. It must be stressed in our model that the individuals
who are suffering the epidemic are not considered to be in infected
state before they gaining the ability to infect others.

Our model is simulated on Erd\"{o}s-R\'{e}nyi (ER) network of
size $N=10^{4}$ with mean degree $\langle k\rangle=10$. In any
case, an individual that get infected at time $t$ will try to
transmit the infection to all its susceptible neighbors at
time $t+1$ (for exactly one time step), then it will lose interest
and never transmit the infection again, i. e., the recover
probability is $\gamma=1.0$. In the simulations, $N_{r}$ (ranging
from $500$ to $2000$) independent realizations are needed.
Moreover, there are two initial conditions: (1) $\rho_{0}$ of
the population are randomly chosen to be in the infected (I)
state as seeds, while the remaining nodes are susceptible; (2)
a proportion $\mu$ of randomly chosen edges in the networks
are regarded as strong edges while the remaining edges are
normal ones.

\section{Theory analysis}
\label{sec:theory}

Based on the edge-based compartmental theory~\cite{weier1,
wang2017unification}, the probability that an individual
owning $k$ connections has received $m$ pieces of pathogen
from its infected neighbors until time $t$ is
\begin{eqnarray}
\phi(k, m, t)=(1-\rho_{0})\dbinom{k}{m}\theta(t)^{k-m}
[1-\theta(t)]^{m},
\label{eq:mphi}
\end{eqnarray}
where $\theta(t)$ denotes the probability at which the epidemic
has not been transmitted through an edge up to time $t$.
Thus the probability that a individual of degree $k$ is in susceptible state is
\begin{eqnarray}
S(k, t)=\sum\limits_{m=0}^{T-1}\phi(k, m, t).
\label{eq:skt}
\end{eqnarray}
Then the fraction of whole susceptible population at time $t$ is
\begin{eqnarray}
S(t)=\sum\limits_{k=1}P(k)S(k, t).
\label{eq:sfraction}
\end{eqnarray}
According to the rules of our model, we obtain
\begin{eqnarray}
\theta(t)=(1-\mu)\theta_{l}(t)+\mu\theta_{h}(t),
\label{eq:theta1}
\end{eqnarray}
where $\theta_{l}(t)$ [$\theta_{h}(t)$] represents the probability
that up to time $t$ the epidemic has not been successfully transmitted
from one end of one normal (strong) edge to the other end. Denoting
$\mu$ as the fractions of strong edges in the networks. Furthermore,
\begin{equation}
\begin{split}
\theta_{l}(t) & =  \xi^{S}_{l}(t)+\xi^{I}_{l}(t)+\xi^{R}_{l}(t), \\
\theta_{h}(t) & =  \xi^{S}_{h}(t)+\xi^{I}_{h}(t)+\xi^{R}_{h}(t),
\end{split}
\label{eq:xihl}
\end{equation}
where $\xi^{y}_{x}(t)$ is the probability that one end of edge
($x=l$ means normal edge while $x=h$ for strong type) is in
the susceptible state while the other end is in state of $y\in
\lbrace S, I, R\rbrace$ up to time $t$.

In a similar way, for a node $\nu$ contacted with a susceptible
neighbor, the probability that $\nu$ has received pathogen $m$
times until time $t$ is
\begin{eqnarray}
\psi(k^{\prime}, m, t)=(1-\rho_{0})\dbinom{k^{\prime}-1}{m}
\theta(t)^{k^{\prime}-m-1}[1-\theta(t)]^{m}.
\label{eq:mpsi}
\end{eqnarray}
Besides, the probability that any edge in the networks contacts
a node of degree $k^{\prime}$ is $k^{\prime}P(k^{\prime})/\langle
k\rangle$ in uncorrelated networks, where $\langle k\rangle$
represents the mean degree of the network. Therefore, we have
\begin{eqnarray}
\xi^{S}_{x}(t)  = \frac{1}{\langle k\rangle}\sum_{k^{\prime}}k^{\prime}
P(k^{\prime})\sum^{T-1}_{m}\psi(k^{\prime}, m, t).
\label{eq:xixs}
\end{eqnarray}
Finally, the dynamics of the system could be tracked by
Eq.~(\ref{eq:xixs}) and the following equations
\begin{equation}
\begin{split}
\frac{d\xi^{R}_{x}(t)}{dt} & =  [1-\lambda(x)]\xi^{I}_{x}(t), \quad \quad x\in \{l,~h\}, \\
\frac{d\theta_{x}(t)}{dt} & =  -\lambda(x)\xi^{I}_{x}(t),  \\
\xi^{R}_{x} & =  \frac{(1-\theta_{x}(t))[1-\lambda(x)]}{\lambda (x)},
\end{split}
\label{eq:dynamics2}
\end{equation}
where $\lambda(x)=\begin{cases}
p, \quad \text{if}\quad x=l,\\
q, \quad \text{if}\quad x=h.
\end{cases}$

Combing Eqs.~(\ref{eq:xixs}) and (\ref{eq:dynamics2}), we obey
\begin{eqnarray}
\hspace{-1.2cm} \frac{d\theta_{x}(t)}{dt}  = \lambda(x)\frac{\sum_{k^{\prime}}k^{\prime}P(k^{\prime})\sum^{T-1}_{m}\psi(k^{\prime}, m, t)}{\langle k\rangle}+1-\lambda(x)-\theta_{x}(t),
\label{eq:dynamics5}
\end{eqnarray}
where $\theta_{h}(t)$ and $\theta_{l}(t)$ can be obtained by
numerically integrating Eq.~(\ref{eq:dynamics5}) with the initial
condition $\theta(0)=1.0$. Then $\theta(t)$ is naturally derived
through Eq.~(\ref{eq:theta1}). We can thus get $S(t)$ by submitting
$\theta(t)$ into Eqs.~(\ref{eq:skt}) and (\ref{eq:sfraction}),
and order parameter $R(\infty)=1-S(\infty)$ as $t\rightarrow \infty $.

Besides, at the end of the spreading process $  d\theta_{x}(t)/dt=0$.
Hence the asymptotic value of $\theta_{x}(\infty)$ obeys
\begin{eqnarray}
\theta_{x}(\infty) & = & \lambda(x)\frac{\sum_{k^{\prime}}k^{\prime}P(k^{\prime})\sum^{T-1}_{m}\psi(k^{\prime}, m, \infty)}{\langle k\rangle}+1-\lambda(x) \\ \notag
				   & = & f_{x}(\theta(\infty)).
\label{eq:thetainfty}
\end{eqnarray}
Combing with Eq.~(\ref{eq:theta1}), we obtain
\begin{eqnarray}
\theta(\infty) = [\mu q+(1-\mu)p][\frac{\sum_{k^{\prime}}k^{\prime}P(k^{\prime})\sum^{T-1}_{m}\psi(k^{\prime}, m, \infty)}{\langle k\rangle}-1]+1,
\label{eq:inftytheta}
\end{eqnarray}
and
\begin{eqnarray}
g(p, \theta(\infty)) = [\mu q+(1-\mu)p][\frac{\sum_{k^{\prime}}k^{\prime}P(k^{\prime})\sum^{T-1}_{m}\psi(k^{\prime}, m, \infty)}{\langle k\rangle}-1]+1-\theta(\infty).
\label{eq:gfunction}
\end{eqnarray}

Next, we further explore the critical behaviors of the
systems under different parameter situations. When $T=1$
and $\rho_{0}\simeq 0$ (i.e., the epidemic is multi-resistant
or even totally-resistant), there exists one solution trivial
$\theta(\infty)=1$ of Eq.~(\ref{eq:inftytheta}). At the
critical point, $f(p, \theta(\infty)$ is tangent to horizontal
axis at $\theta(\infty)=1$. Accordingly,
we could obtain the continuous critical condition of the model as
\begin{eqnarray}
\frac{dg(p, \theta(\infty))}{d\theta(\infty)}\Big
|_{\theta(\infty)=1}=0.
\label{eq:deltag}
\end{eqnarray}
Based on Eq.~(\ref{eq:deltag}), we can elicit the following relationship
\begin{eqnarray}
[\mu q+(1-\mu)p]\frac{\langle k^{2}\rangle -\langle k\rangle}
{\langle k \rangle}=1.
\label{eq:thresholdrelationship}
\end{eqnarray}
Naturally, the critical transmission probability can be calculated.
In detail, the critical point $p_{c}^{\rm II}=\langle k \rangle/(\langle k^{2}
\rangle -\langle k\rangle)$ if $p=q$, being in agreement with the
threshold of the classical SIR spreading model~\cite{claudiothreshold}.
While there are two cases to be considered for the more widely existed
relationship $p\neq q$ in reality
\begin{equation}
\begin{split}
& p_{c}^{\rm II}~\text{or}~q_{c}^{\rm II}  = \frac{\langle k \rangle}{\langle k^{2}\rangle -\langle k\rangle},~\text{if}~\mu=0~\text{or}~1,\\
& p_{c}^{\rm II} = \frac{\langle k \rangle}{(\langle k^{2}\rangle -\langle k\rangle)-(1-\mu)}-\frac{\mu q}{1-\mu},~\text{otherwise}.
\label{eq:t1threshold}
\end{split}
\end{equation}
According to Eq.~(\ref{eq:thresholdrelationship}), we could further
expected that null threshold $p_{c}^{\rm II}=0$ would arises when
$\mu\geq \mu_{c}^{\rm II}= -b+\sqrt{b^{2}+4q\langle k\rangle}/2q$, where
$b=(\langle k^{2}\rangle -\langle k\rangle-1)q$.

\begin{figure}[!h]
\includegraphics[width=\textwidth]{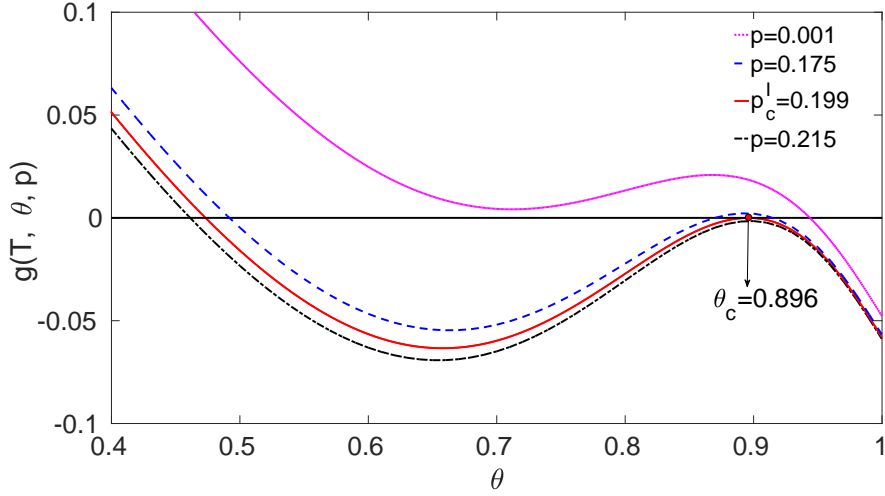}
\caption{Illustration of graphical solutions of Eq.~(\ref{eq:gfunction})
for $T=3$. The black solid line is the horizontal axis, and
the red dot denotes critical value $\theta_{c}$. Other parameter is taken as $\rho_{0}=0.1$ and $\mu=0.477$.
}
\label{fig:gfunction}
\end{figure}
For the case of $T>1$ and $\rho_{0}\simeq 0$ (i.e., the epidemic
owns partial drug-resistance), $\theta(\infty)$
is still the trivial solution of Eq.~(\ref{eq:gfunction}),
but $f(\theta(\infty))$ will not be tangent to horizontal axis
at the point $\theta(\infty)=1.0$. With increase of $\rho_{0}$,
$\theta(\infty)=1.0$ is not a solution of the Eq.~(\ref{eq:thetainfty})
any more. If Eq.~(\ref{eq:thetainfty}) has two stable fixed points,
only the largest one is physical meaningful since $\theta(t)$
decreases with $t$ from $1$. As shown in Fig.~\ref{fig:gfunction},
the physical meaningful solution of Eq.~(\ref{eq:thetainfty}) is
the only one stable fixed point when $p$ is small. At the critical
point $p_c^{\rm I}$, the physical meaningful solution of
Eq.~(\ref{eq:thetainfty}) is the tangent point. To determine the
value of $p_c^{\rm I}$, we can numerically solve Eq.~(\ref{eq:thetainfty})
and
\begin{equation}\label{cri}
\frac{dg(p, \theta(\infty))}{d\theta(\infty)}|_{\theta_c(\infty)}=0,
\end{equation}
where $\theta_c(\infty)$ is the critical value that the infection
has not transmitted through an edge. We can further calculate the
final expression of critical transmission probability
\begin{eqnarray}
p_{c}^{\rm I}=\frac{H(\theta_{c}, \rho_{0}, T, \langle k\rangle)-\mu q}{1-\mu},
\label{eq:finalthreshold}
\end{eqnarray}
where $H(\theta_{c}, \rho_{0}, T, \langle k\rangle)  = \langle k\rangle/(1-\rho_{0})\sum\limits_{k}kP(k)\sum\limits_{m}^{T-1}
\theta_{c}^{k-m-2}(1-\theta_{c})^{m-1}[(k-1)(1-\theta_{c})-m]$.
From Eq.~(\ref{eq:finalthreshold}), we know that
$p_c^{\rm I}$ is correlated with $\mu$, $q$, $T$, $\rho_0$, and
$\langle k\rangle$. When $p>p_c^{\rm I}$, we find that
Eq.~(\ref{eq:thetainfty}) has only one small stable fixed point.
From the above statements, the physical meaningful solution of
Eq.~(\ref{eq:thetainfty}) decreases discontinuously from a large
value to a small one. From the perspective of nonlinear dynamics,
the system exhibits a bifurcation phenomenon, which indicates
$\theta(\infty)$ decreases discontinuously with $p$, that is
$R(\infty)$ increases discontinuously with $p$. If Eq.~(\ref{eq:thetainfty})
has only one stable fixed point, $R(\infty)$ increases
continuously with $p$. The critical point that the growth
pattern of $R(\infty)$ changes can be determined by
simultaneously solving Eqs.~(\ref{eq:thetainfty}), (\ref{cri}) and
\begin{equation}
\begin{split}
\frac{d^{2}g(p, \theta_{s}(\infty))}{d\theta_{s}(\infty)^{2}}\Big |_{\theta _{s}(\infty)} =0.	
\label{eq:deltag2}
\end{split}
\end{equation}

\section{Results}
\label{sec:results}

In this section, we perform extensive numerical simulations
for our proposed model with totally ($T=1$) and partial
($T>1$) drug-resistant respectively.

\subsection{Totally drug-resistant}
\label{subsec:t1}

We first study the model with totally drug-resistant, i.e.,
$T=1$. In this situation, individuals would easily get infected
if they receive the epidemic from the neighbors for one time. This case
corresponds to the diffusion of an epidemic that is difficult to treat, or with total drug
resistance; where our model is equivalent to the classical
SIR model, but with two types of connections along which
epidemics spread at different probability.

\begin{figure}
\includegraphics[width=\textwidth]{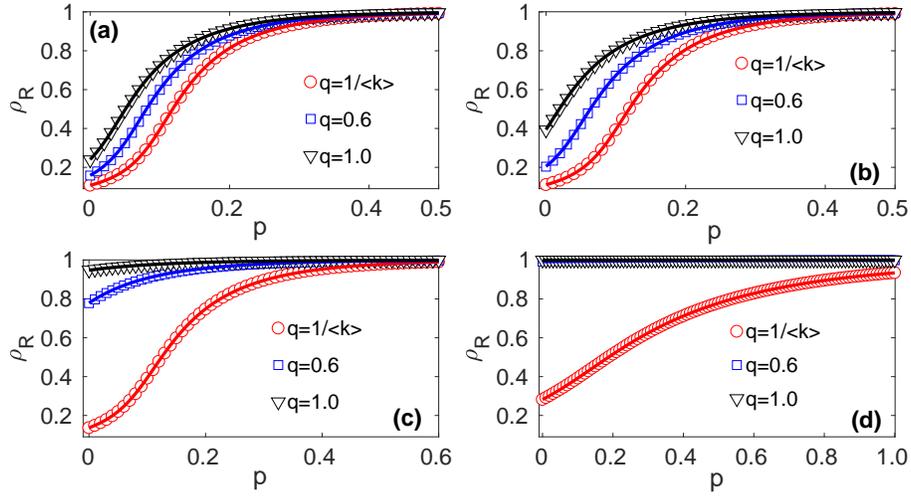}
\caption{The final steady density of removed individuals versus $p$ for four different values of raised infection probability $q$ of each $\mu$, where four four different fractions of edges are initially set as strong ones: (a) $\mu=0.07$, (b) $\mu=1/\langle k\rangle=0.1$, (c) $\mu=0.3$ and (d) $\mu=0.8$. The results are averaged over $N_{r}=1000$ independent realizations. The other parameter is taken as $\rho_{0}=0.1$.
}
\label{fig:t1mufunctionq}
\end{figure}

It is obvious in Fig.~\ref{fig:t1mufunctionq} that the
phase transitions for case $q<1.0$ and $\mu\leq1/\langle k
\rangle$ and case $q<  1/\langle k\rangle$ and $\mu>1/
\langle k\rangle$ belong to continuous class. While null
threshold (exhibit the lack of an epidemic threshold and
always show a finite fraction of infected population)
~\cite{nullthreshold} occurs if and only if $q>1/k$
for $\mu>1/\langle k\rangle$. For continuous transitions,
the threshold decrease with increasing $q$, and the fractions
of strong edges $\mu$ in the networks. In contrast to
the case $q>1/\langle k\rangle$, the networks with a large
number of strong edges actually instead suppress the epidemic
spread when $q<p_{c}^{\rm II}$. Based on above results,
there exists a threshold of raised transmission probability
$q_{c}^{\rm II}=1/\langle k\rangle$ predicted from the bond percolation~\cite{op1,op2,op3}, beyond which strong edges
are considered to be strong enough to facilitate the epidemic
spread, by speeding up the transmissions of pathogen along them.

%\begin{figure}[!ht]
%\includegraphics[width=0.5\textwidth]{t1_pfunctionq.eps}
%\caption{The density of recovered individuals in the final state is plotted versus $q$ for three different values of $p$ for each value of $\mu$ on Erd\H os-R\'{e}nyi ER networks with $\langle k\rangle=6$. The results are averaged over $N_{r}=1000$ independent realizations.
%}
%\label{fig:t1_pfunctionq}
%\end{figure}
%Fig.~\ref{fig:t1_pfunctionq} shows that continuous transitions only occur for $p>p_{c}=\frac{1}{\langle k\rangle}$, independent of $\mu$.

\begin{figure}
\includegraphics[width=0.55\textwidth]{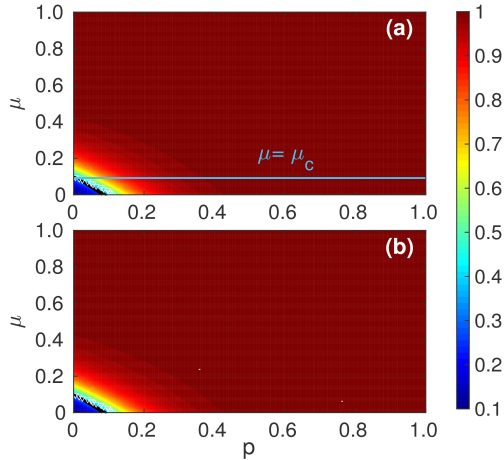}
\caption{The final stationary distributions of removed individuals as function of $p$ and $\mu$ for $T=1$, where the final results are respectively obtained from (a) simulations and (b) numerical integration of Eq.~(\ref{eq:dynamics5}), combing with Eqs.~(\ref{eq:xixs}) and (\ref{eq:dynamics2}). The numerically obtained critical values of transmission probability $p$  are illustrated as white inverted triangles to enable a comparison with the solid black line from theoretical prediction based on Eq.~\ref{eq:t1threshold}. The results are obtained by averaging over $N_{r}=1000$ independent realizations. Moreover, the light blue solid line give the critical value of $\mu_{c}=(-b+\sqrt{b^{2}+4q\langle k\rangle})/(2q)$, where $b=(\langle k^{2}\rangle -\langle k\rangle-1)q$, beyond which the null threshold of outbreak begins to emerge. The other parameter are taken as $q=1.0$ and $\rho_{0}=0.1$.
}
\label{fig:t1}
\end{figure}

The theoretical value of outbreak threshold for $T=1$ could
be calculated from Eq.~(\ref{eq:t1threshold}), so one can make
a comparison with the numerical critical value corresponding
to the position of susceptibility peak~\cite{claudiothreshold}.
The analytical threshold by means of edge-based compartmental
theory shows a good agreement with the numerically evaluated
threshold. The results illustrated in Fig.~\ref{fig:t1mufunctionq}
and Fig.~\ref{fig:t1} prove that the edge-based compartmental
theory provides hence a good tool to portray the dynamics of
our model in theory, with a high accuracy.

As expected, Fig.~\ref{fig:t1} implies that lack of effective treatment or serious disease of totally drug-resistant could facilitate the outbreak of epidemics because individuals are doomed to be infected if they contact the patients with high frequency and receive the pathogen only once. Fig.~\ref{fig:t1mufunctionq} shows that the transitions are continuous for $T=1$, which is further supported by the smooth color bitmap transitions for $\mu<\mu_{c}$ showed in Fig.~\ref{fig:t1}. As expected, the boundary illustrated in Fig.~\ref{fig:t1} reveals the favored role of strong edges to outbreaks of epidemic. Besides, both numerical simulations and theoretical analysis manifest the existence of null threshold when there are more than $\frac{N\langle k\rangle\mu_c}{2}$ strong edges in the network. This indicates that a small number of infected individuals ($\rho_{0}$ of the whole population) could still spread the totally drug-resistant epidemic out to capture the whole population, merely relying on few ($\mu_{c}\simeq 10^{-2}$) strong contacts of the network. On the other hand, continuous transitions reveals that a slow growth of the epidemic activity for increasing spreading rates makes epidemic less threatening, which could leave us enough time to take measures for the treatment and control of these infectious diseases.
%%%%%%%%%%%%%%%%%%%%%%%%%%%%%%%%%%%%%%%%%%%%%%%%%%%%%%%%%%%%%%%%%%%%%%%%%%%%%%%%%%%%%%%%%%%%%%%%%%%%%%%%%%%%%%%%%%%%%%%%%%%%%%%%%%%%%%%%%%%%%%%
%%%%%%%%%%%%%%%%%%%%%%%%%%%%%%%%%%%%%%%%%%%%%%%%%%%%%%%%%%%%%%%%%%%%%%%%%%%%%%%%%%%%%%%%%%%%%%%%%%%%%%%%%%%%%%%%%%%%%%%%%%%%%%%%%%%%%%%%%%%%%%%
\subsection{Partial drug-resistant}
\label{subsec:t23}
We further study the situations that epidemic is partial drug-resistant,
i.e., $T>1$. A larger value of $T$ requires the individual
to be exposed with more infection from distinct neighbors to affirm
the attack ability of the epidemic. The model for $T>1$ is assumed
to describe a society where good medical treatment especially
antibiotic therapy is available for the individuals once the
epidemic is diagnosed.

\begin{figure}[!ht]
\includegraphics[width=\textwidth]{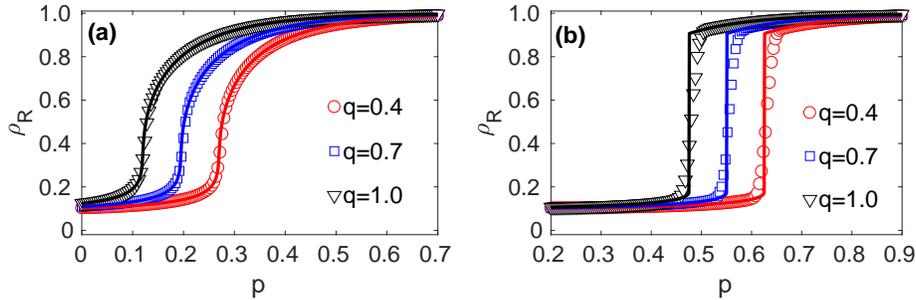}
\caption{The final density of removed individuals versus $p$ for three different values of raised infection probability $q$. In detail, $T=2$ in (a) while $T=3$ in (b). $\mu=0.25$ of the total edges are initially set as strong edges while the others are normal ones. The simulation results represented by open markers are obtained by averaging $N_{r}=2000$ independent realizations, in comparison with the corresponding solid lines from analytical predictions based on Eq.~(\ref{eq:dynamics5}).
}
\label{fig:mu_functionqt23}
\end{figure}

We firstly present both the simulation results and numerical
integration of Eq.~(\ref{eq:dynamics5}) in Fig.~\ref{fig:mu_functionqt23}
to investigate how the spread of epidemics depend on the
strength of the strong edges (value of $q$) and infection
threshold $T$. We find that the edge-based compartmental
theory also gives accurate estimates of the size of epidemic
spread on ER networks for $T>1$. Besides, it is found that
stronger edges of higher $q$ could impel the outbreak of
epidemics to take place, which is in accordance with what
we have observed in Fig.~\ref{fig:t1}. Sharp jumps of infection
sizes from numerical simulations and theoretical predictions
presented in Fig.~\ref{fig:mu_functionqt23} indicate
discontinuous transitions. Furthermore, more sharp
transitions could be observed for higher $T$.
The type of transition can be determined by using
bifurcation analysis of Eq.~(\ref{eq:gfunction}).

The occurrence of discontinuous transitions with $T>1$
could be attributed to occurrence of avalanche outbreaks
as $p$ is beyond the threshold~\cite{avalanche1,avalanche2}.
In the present model, one consider that the spread dynamics
of the epidemics start with a small number $\rho_{0}$ of
nodes which have initially received the pathogen and become
infected, regardless of the infection threshold $T$. In the
following, some of neighbors of these initial seeds, due to
the incoming pathogen brought about by enough high infection
probability $p\geq p_{c}$, may receive the pathogen for more
than $T$ times; next, these newly fully infected sites also
spread pathogen, during the same time step, again to their
own neighbors which have received the pathogen below threshold
$T$. Although large number of the left nodes are still
susceptible, they have received the pathogen for some times;
which means that only one or two more arrival of pathogen
would trigger more new infections to make all of them infected.
In other words, these susceptible nodes are in critical state.
The process can go on and lasts until none of the neighbor nodes
goes above the infection threshold, thus this avalanche stops;
which showing a massive expansion of infection population in
the networks. Moreover, in comparison with the case $T=2$,
larger number of susceptible nodes are in critical state for
$T=3$, leading to a more drastic avalanche which indicates a
more obvious discontinuous transition (see Fig.~\ref{fig:mu_functionqt23}).
Since the susceptible nodes in critical state could be in one
of two critical states for $T=3$: have received the pathogen
either once or twice; rather than only one critical state for
$T=2$. Meanwhile thresholds of outbreaks are greatly increased
for $T=3$ (see Fig.~\ref{fig:mu_functionqt23}), owing to the
fact that larger infection probability $p$ are needed to drive
the pathogen to reach more susceptible areas to make the nodes
there be in critical states in the avalanche process. Overall,
a necessary condition for discontinuous transitions is an
abundance of susceptible nodes in the critical state on the
underlying network.

\begin{figure}[!ht]
\includegraphics[width=\textwidth]{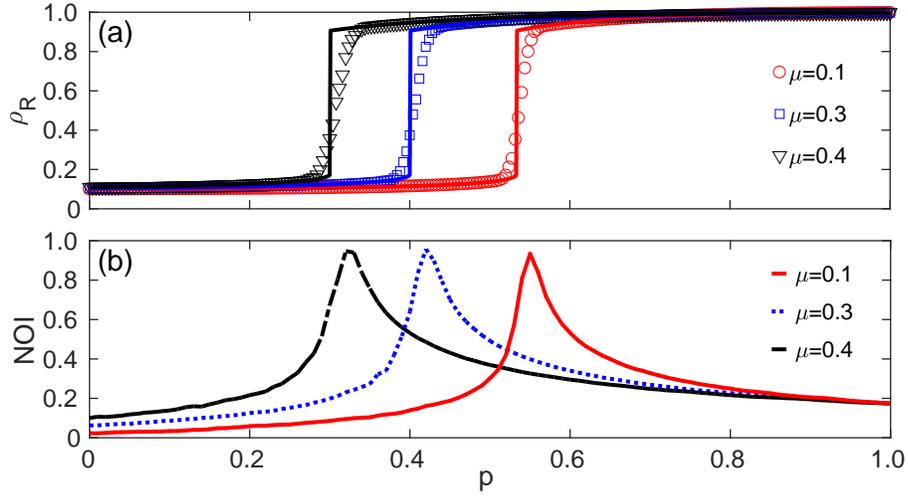}
\caption{(a) The final density of removed individuals versus
$p$ for three different values of $\mu$, where both simulation
results (open markers) and theoretical predictions (solid lines)
are plotted to make a comparison. (b) Simulation results of NOI
(number of iterations) as a function of $p$ with three difference
values of $\mu$. The raised transmission probability is $q=1.0$,
and $T=3$. The results are obtained by  averaging $N_{r}=2000$
independent realizations.
}
\label{fig:t3_difmu_both}
\end{figure}

The theoretical value of $p_{c}$ for $T>1$ can be calculated
from Eqs.~(\ref{eq:deltag}) and (\ref{eq:deltag2}), while the
numerical critical value can be estimated by identifying the
number of iterations (denoted as NOI) at which at least one
individual get infected~\cite{weier1,noi1}. In Fig.~\ref{fig:t3_difmu_both},
it could be observed that the NOI exhibits a peak of maximum
value; which shows a remarkable agreement between theory and
numerics in terms of the quantities $\rho_{R}$ and $p_{c}^{\rm I}$.
This provides further proof of the correctness of analysis based
on edge-based compartmental theory. Also, Fig.~\ref{fig:t3_difmu_both}
provides a important evidence both in numerical simulations and
in theory about the role of strong contacts in governing spread of
epidemics, that is, more strong edges could not only facilitate
the outbreaks of epidemics by reducing outbreak threshold but
also promote the spread of epidemics (because more strong edges
in the networks means higher equivalent infection probability
of epidemics). Again, the integration of analytical equations
not only predicts the position of the threshold, but also identifies
discontinuous types of the transitions.

\begin{figure}[!ht]
\includegraphics[width=\textwidth]{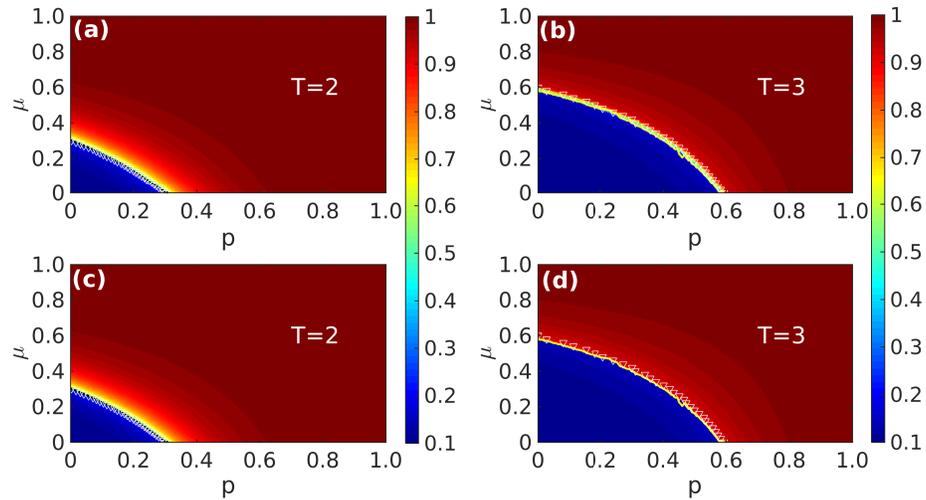}
\caption{The final stationary distributions of removed individuals as function of $p$ and $\mu$ for $T=2$ (a)(c) and $T=3$ (b)(d), where the final results are respectively obtained from simulations (shown in (a)(b)) and numerical integration of Eqs.~(\ref{eq:dynamics5}), (\ref{eq:xixs}) and (\ref{eq:dynamics2}),
as shown in (c-d). The numerically obtained critical values of
transmission probability $p$ through NOI method are illustrated as
white inverted triangles to enable a comparison with the solid black
(for $T=2$) or yellow (for $T=3$) line from theoretical prediction
based on Eq.~(\ref{eq:t1threshold}). The results are obtained by
averaging over $N_{r}=1000$ independent realizations. The other
parameter are taken as $q=1.0$ and $\rho_{0}=0.1$.
}
\label{fig:t23}
\end{figure}

Fig.~\ref{fig:t23} presents a comprehensive review about the
dependence of the infection size on both fractions of strong
edges $\mu$ and infection probability $p$ for $T>1$; containing
numerical simulations and theoretical estimations. It could be
found that simulation results [see Figs.~\ref{fig:t23}(a)
and (b)] and numerical integration [see Figs.~\ref{fig:t23}(c)
and (d)] show a good accordance with each other. In the same
figure, the thresholds obtained from theoretical predictions
are also plotted to make a comparison with numerical estimation
of NOI, revealing a rather satisfactory agreement. As the
infection threshold grows from $T=2$ to $T=3$, both $\mu_{c}$
and position of the threshold $p_{c}$ are greatly increased;
however, the nature of the transition remains unchanged. In
addition, despite of infection threshold $T$, the emergence
of null threshold with $\mu>\mu_{c}$ tells us that large
number of strong contacts in the society could make the
prevalence of pathogens critical easy all the time by
forming sizable cluster (larger than giant cluster of the
network because $\mu_{c}>1/\langle k\rangle$) which could
reach most of the nodes, without help of the transmissions
along normal edges. Notice that the dependence of the
threshold boundary (indicated by both solid lines and
inverted triangles) is another piece of evidence in favor
of the understand of what roles strong edges, infection
probability and infection threshold play in favoring
epidemic diffusion. In detail, compared with $T=2$
[Figs.~\ref{fig:t23}(a) and (c)], more edges even the
majority of total edges (nearly 60\%) when $p=0$, are
needed to be strong for $T=3$ [Figs.~\ref{fig:t23}(b) and
(d)] so that the whole population are occupied by the
epidemic. This means that high infection threshold $T$
indeed delay the outbreak of epidemics and emergence
of null threshold, in contrast with the acceleration
of strong edges on the spread of epidemic. This is
what is observed in Fig.~\ref{fig:t23}.

\subsection{Roles of initial seeds}

\label{subsec:seed}
\begin{figure}
\includegraphics[width=\textwidth]{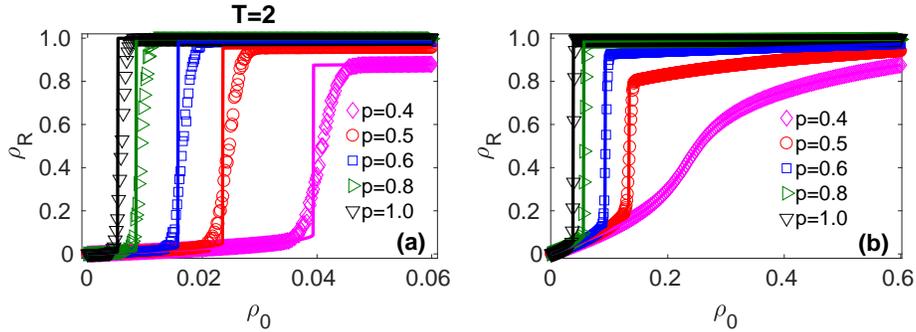}
\caption{The final density of removed individuals as function
of $\rho_{0}$ for five different values of $p$. In detail, $T=2$
in (a) while $T=3$ in (b). $\mu=0.01$ of the total edges are
initially set as strong edges while the others are normal ones.
The simulation results represented by open markers are obtained
by averaging $N_{r}=2000$ independent realizations, in comparison
with the corresponding solid lines from theoretical predictions
based on Eq.~(\ref{eq:dynamics5}). The other parameter is taken
as $q=1.0$.
}
\label{fig:museed}
\end{figure}

In this subsection, we turn to elucidating the effect of
number of initial seeds on epidemic spread dynamics. We
note that the transitions shown in Fig.~\ref{fig:museed}
are discontinuous when infection probability $p$ is large
enough. Furthermore, Fig.~\ref{fig:museed} displays a
degree of independence of infection sizes on number of
initial seeds as $\rho_{0}\geq 0.1$. Since only 0.01 of
total edges are assumed to be normal ones in
Fig.~\ref{fig:museed}. By combing with the illustration
in Fig.~\ref{fig:t3_difmu_both}, we can deduce that the
outbreaks and prevalence of epidemics are more relatively
independent of the number of initial seeds as $\rho_{0}$
getting larger than 0.1 for higher fractions of strong
edges. This is why we adopt $\rho_{0}=0.1$ throughout
this paper. It is also interesting that the epidemic
does not spread out immediately even if the individuals
would inevitably received the pathogen on condition that
their neighbors get infected, i.e., $p=1.0$. The reason
for this is that one susceptible individual would probably
get infected only when more than one neighbors are in the
infected state. This is to say there exists a critical
value of $\rho_{0}$ above which initial seeds could promote
the outbreak of epidemics.

\section{Conclusion}
\label{sec:conclusion}
In this paper we have proposed a biologic spread model considering not only drug resistance  curved by infection threshold $T$, but also heterogeneous contributions of connections by simply classing the edges of the networks into two classes: normal edges and strong ones. Then we have performed both numerical simulations and theoretical analysis employing edge-based compartmental theory to deeply investigate the system behaviors. In detail, we have firstly explored the spread dynamics of our model for $T=1$. Furthermore, we deduce the kinetic equations to curve the spread dynamics to allow us to give a precise prediction of final infection size. By means of edge-based compartmental theory, we also have obtained the expressions for the threshold of continuous transitions, revealing a good agreement with the numerical estimations. Overall, both simulations and theoretical analysis show two main results: (1)for $\mu<\mu_{c}$ the system gives a evidence for a continuous transition which is vulnerable to the possibility of outbreaks (low threshold); (2) while abundant strong edges themselves could promote the outbreaks and prevalence of epidemics, i.e., null threshold begin to emerge as $\mu>\mu_{c}$. Overall, the results for $T=1$ suggest that the populations are vulnerable to the epidemics with total resistance. Whereas continuous transition indicates a slow growth of the epidemic activity that makes epidemic less threatening, leaving us enough time to take measures for the treatment and control of epidemic.

In the following we extend our study to the situations of $T>1$ to map the spread of epidemic with partial drug resistance. By means of edge-based compartmental theory, we have further developed a method to find the positions of thresholds for discontinuous transitions, as well as an exact expression for the epidemic threshold. Both numerical simulations and analytical predictions show that the infection size depends on the fractions of strong edges of network $\mu$, the strength of strong edges $q$ and infection threshold $T$. The thresholds obtained from theoretical approach are found be in good accordance with numerical estimations of NOI. Differing from the case $T=1$, we find a confirmation that the discontinuous transition occurs when sufficient initial infected nodes are seeded in the network and the strength of strong contacts is high enough, but at a more larger threshold. Meanwhile we have given the interpretation about the origin of the discontinuous epidemic transition occurring in this kind of system for large $T$, that is occurrence of avalanche outbreak~\cite{avalanche1,avalanche2} brought about by large number of susceptible individuals in critical state. Taken together with the results from agent-based simulations and the theoretical predictions, we have concluded that both the strength of strong contacts $q$ and the fractions of strong edges $\mu$ play important positive roles in promoting the outbreak and prevalence of epidemics with drug resistance. Nevertheless higher $T$ corresponding to sophisticated  antibiotic resources or more simply accessible specific medicines actually hinder or delay the coming of epidemic prevalence; but that if the strong contacts are not abundant enough the true, inhibiting effect of the $T$ becomes evident.

It is also worth noticing that our theoretical model does not take more realistic complex mechanisms such as dynamic changes of infection threshold, individual variation in resisting different epidemics owning drug resistance into account. Also, the correlation between strong contacts and drug resistance has not been considered in the model because of lack of empirical data. However, these findings of our study lead directly to two important suggestions for a logical treatment of epidemics: (1) imposing contact restrictions on connections especially strong contacts to inhibit the spread such as the shutdown of individual connections or isolation of cities,  which has gained wide acceptance~\cite{contact1,control1,contact2}; (2) keeping persistent development of new wonder medicines or new therapies (such as medicine cocktail therapy) to sustain high values of infection threshold $T$. Of course, sustainable research inputs for the development are indispensable.

What is more important this study has revealed is that, despite of enough and good medical resources to enable members of developed society repel the epidemics more than one time, sharp massive outbreaks of epidemics resulting from drug resistances mutations seems to be still a possible serious threat. Even worse, economic recession in developed regions would greatly aggravate contagion of epidemics which have developed partial or even complete resistance to antibiotics; leaving people in a more dangerous situation.

Drug resistance and contact connections are two very important ingredients for biological spreading processes. The investigation of epidemic diffusion with the two mechanisms in general complex contagion situations remains a very important and interesting avenue for future research activity. Our approach may be considered a basic reference framework for the further description and exploration of epidemics spread with other ingredients, in addition to drug resistance and heterogeneous contacts.
\section*{Acknowledgments}
This work was supported by China Postdoctoral Science Foundation (CPSF) Grant No. 2015M582532 and by China Scholarship Council (CSC), Grant No. 201606075021.
%\section*{Appendix A: Extended pair approximation}
%\label{apa}
%
%
%\section*{Appendix B: Analysis for strategy phases based on spatial replicator equation}
%\label{apb}

\end{document}